\documentclass[12pt]{article}

\pdfoutput=1 
\usepackage{graphicx} 
\usepackage{color}                                                                                                                                                                                                                                                                                                                                                                                                                                                                                                                                                                                                                                                                                                                                                                                                                                                                                                                                                                                                                        
\usepackage{hyperref}                                                                                                                                                                                                                                                                                                                                                                                                                                                                                                                                                                                                                                                                                                                                                                                                                                                                                                                                                                                                                                                                                                                                                                                                                                                                                                                                                                                                                                                                                                                                                                                                                                                                                                                                                                                                                                                                                                                                                                                                                                                                                                                                                                                                                                                                                                                                                                                                                                                                                                                                                                                                                                                                                                                                                                                                                                                                                                                                                                                                                                                                                                                                                                                                                                                                                                                                                                                                                                                                                                                                                                                                                                                                                                                                                                                                                                                                                                                                                                                                                                                                                                                                                                                                                                          
\usepackage{cancel,tabularx,moreverb,fancybox,amsmath,float,bm,braket,slashbox,txfonts,amssymb,bm,accents}
\usepackage[top=30truemm,bottom=30truemm,left=25truemm,right=25truemm]{geometry}
\usepackage{latexsym}
\usepackage{here}

\newcommand{\ex}[1]{\mathrm{e}^{#1}}

\newcommand{\pa}[1]{\left(#1 \right)}

\newcommand{\br}[1]{\left[#1 \right]}

\newcommand{\bb}[1]{\mathbb{#1}}

\newcommand{\ca}[1]{\mathcal{#1}}

\newcommand{\abs}[1]{\left|#1\right|}

\newcommand{\ar}[1]{\xrightarrow[#1]{}}

\newcommand{\ti}[1]{\tilde{#1}}

\newcommand{\fr}{\frac}
\newcommand{\s}[1]{\sqrt{#1}}

\def\be{\begin{equation}}
\def\ee{\end{equation}}
\def\ba{\begin{eqnarray}}
\def\ea{\end{eqnarray}}

\def\del{{\partial}}

\def\la{{\lambda}}
 \def\w{{\omega}}

 \def\d{{\delta}}

 \def\a{{\alpha}}
 
 \def\l{{\lambda}}
 \def\G{{\Gamma}}

 \def\e{{\epsilon}}

 \def\p{\partial}

\def\tr{{\text{tr}}}

\def\dd{{\mathrm{d}}}

\def\ek{{\eta_{2+3+\kappa}}}

  \makeatletter
    
    \@addtoreset{equation}{section}
  \makeatother

\begin{document}

\begin{titlepage}
\thispagestyle{empty}

\begin{flushright}
YITP-1863
\\

\end{flushright}

\bigskip

\begin{center}
\noindent{{\large \textbf{
Large $c$ Virasoro Blocks from Monodromy Method\\
beyond Known Limits
}}}\\
\vspace{2cm}
Yuya Kusuki
\vspace{1cm}

{\it
Center for Gravitational Physics, \\
Yukawa Institute for Theoretical Physics (YITP), Kyoto University, \\
Kitashirakawa Oiwakecho, Sakyo-ku, Kyoto 606-8502, Japan.
}
\vskip 2em
\end{center}

\begin{abstract}
In this paper, we study large $c$ Virasoro blocks by using the Zamolodchikov monodromy method beyond its known limits. 
We give an analytic proof of our recent conjecture \cite{Kusuki2018, Kusuki2018b}, which implied that the asymptotics of the large $c$ conformal blocks can be expressed in very simple forms, even if outside its known limits, namely the semiclassical limit or the heavy-light limit. In particular, we analytically discuss the fact that the asymptotic behavior of large $c$ conformal blocks drastically changes when the dimensions of external primary states reach the value $\frac{c}{32}$, which is conjectured by our numerical studies. The results presented in this work imply that the general solutions to the Zamolodchikov recursion relation are given by Cardy-like formula, which is an important conclusion that can be numerically drawn from our recent work \cite{Kusuki2018, Kusuki2018b}.
Mathematical derivations and analytical results imply that, in the bulk, the collision behavior between two heavy particles may undergo a remarkable transition associated with their masses.
\end{abstract}
 
\end{titlepage}

\restoregeometry

\tableofcontents
\section{Introduction}

Conformal Field Theories (CFTs) have brought about great breakthroughs for theoretical physics. In particular, if one focuses on two-dimensional CFTs, one realizes that CFTs are highly constrained by the Virasoro symmetry and, therefore, in two-dimensions, CFTs are specified by only a central charge, a spectrum of primary states, and their OPE coefficients. Once these data are known, one can evaluate correlators on any Riemann surface without boundaries. Moreover, such CFT data are constrained by the crossing symmetry, which leads to the conformal bootstrap equation \cite{Belavin1984,Ferrara1973,Polyakov1974,Rattazzi2008}. Recently, CFTs have been attracting much attention in the scientific community as tools to probe AdS gravities in the context of the AdS/CFT duality. 

For both the conformal bootstrap and the AdS/CFT duality, conformal blocks provide important contributions \cite{Heemskerk2009,Heemskerk2010,El-Showk2012,Fitzpatrick2013,Fitzpatrick2013a,Hijano2016}. In particular, in AdS$_3$/CFT$_2$, the semiclassical Virasoro blocks have been used to probe information loss, which appears in CFT$_2$ as forbidden singularities and exponential decay at late times \cite{Fitzpatrick2014,Fitzpatrick2015,Fitzpatrick2016,Fitzpatrick2017}. Furthermore, those semi-classical blocks can be computed in the dual AdS$_3$ gravity  \cite{Alkalaev2015,Hijano2015,Hijano2015a,Alkalaev2016}, which means that the conformal blocks themselves have some gravity interpretation. Some other progress attributed to conformal blocks include the study of the dynamics of the Renyi entropy \cite{Caputa2015, Asplund2015,Kusuki2018} and out-of-time-ordered correlators (OTOCs)  \cite{Roberts2015}. (See also \cite{Alkalaev2016b, Chen2016,Alkalaev2016a, Alkalaev2017, Maxfield2017,Alkalaev2018,Hikida2018c,Banerjee2018,Hikida2018a} for further developments in this direction.)

Although there have been a lot of studies about conformal blocks, we have not reached a perfect understanding of them. Even if we restrict ourselves to large $c$ CFTs, there is no closed expression for the conformal blocks, except for the case of special limits. Nevertheless, in our recent works \cite{Kusuki2018,Kusuki2018b}, we numerically find simple expressions for large $c$ conformal blocks  by using the Zamolodchikov recursion relation \cite{Zamolodchikov1987,Zamolodchikov1984}. We particularly study
the Virasoro block for the correlator $\braket{O_B(\infty)O_B(1)O_A(x)O_A(0)}$ in the $O_A(x)O_A(0)$ OPE channel, in which 

\newsavebox{\boxpa}
\sbox{\boxpa}{\includegraphics[width=160pt]{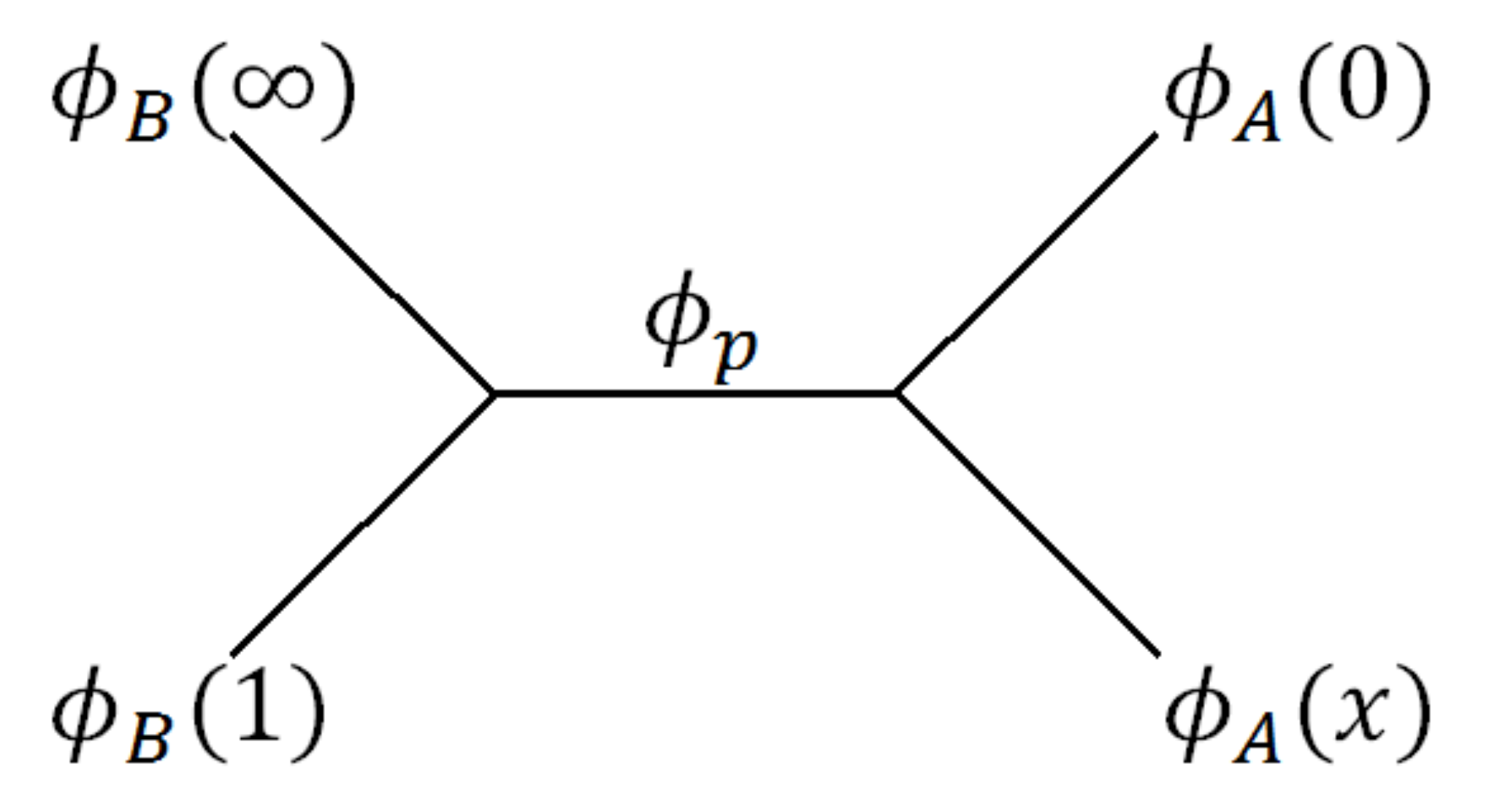}}
\newlength{\paw}
\settowidth{\paw}{\usebox{\boxpa}} 

\begin{equation*}
 \ca{F}^{AA}_{BB}(h_p|x) \equiv \parbox{\paw}{\usebox{\boxpa}},
\end{equation*}
which we call {\it AABB blocks}, and work out the asymptotic behavior of the large $c$ blocks,
\begin{enumerate}
\item light-light region: $h_A, h_B<\fr{c}{32}$ 
\begin{equation}\label{eq:pre4}
\begin{aligned}
&\log \ca{F}^{AA}_{BB}(h_p|x)\\
&\ar{x \to 1}
-\fr{c-1}{12}\pa{1-\s{1-\fr{24}{c-1}h_A}}\pa{1-\s{1-\fr{24}{c-1}h_B}}\log \pa{1-x}               +    O(\log\log(1-x))
\end{aligned}
\end{equation}

\item heavy-heavy region: $h_A, h_B>\fr{c}{32}$ 
\begin{equation}\label{eq:pre5}
\log \ca{F}^{AA}_{BB}(h_p|x)\ar{x \to 1}
\pa{\fr{c-1}{24}-h_A-h_B}\log \pa{1-x}       +    O(\log\log(1-x)).
\end{equation}
\end{enumerate}
Note that in the heavy-light region ($h_A>\fr{c}{32}, h_B<\fr{c}{32}$ or $h_A<\fr{c}{32}, h_B>\fr{c}{32}$), we cannot obtain the asymptotics of the conformal blocks from our recent results \cite{Kusuki2018,Kusuki2018b} due to technical difficulties.

However, for {\it ABBA blocks} defined below, there is no difficulty in the evaluation of asymptotics, and we can obtain them for the conformal blocks in the whole region,
\newsavebox{\boxpd}
\sbox{\boxpd}{\includegraphics[width=160pt]{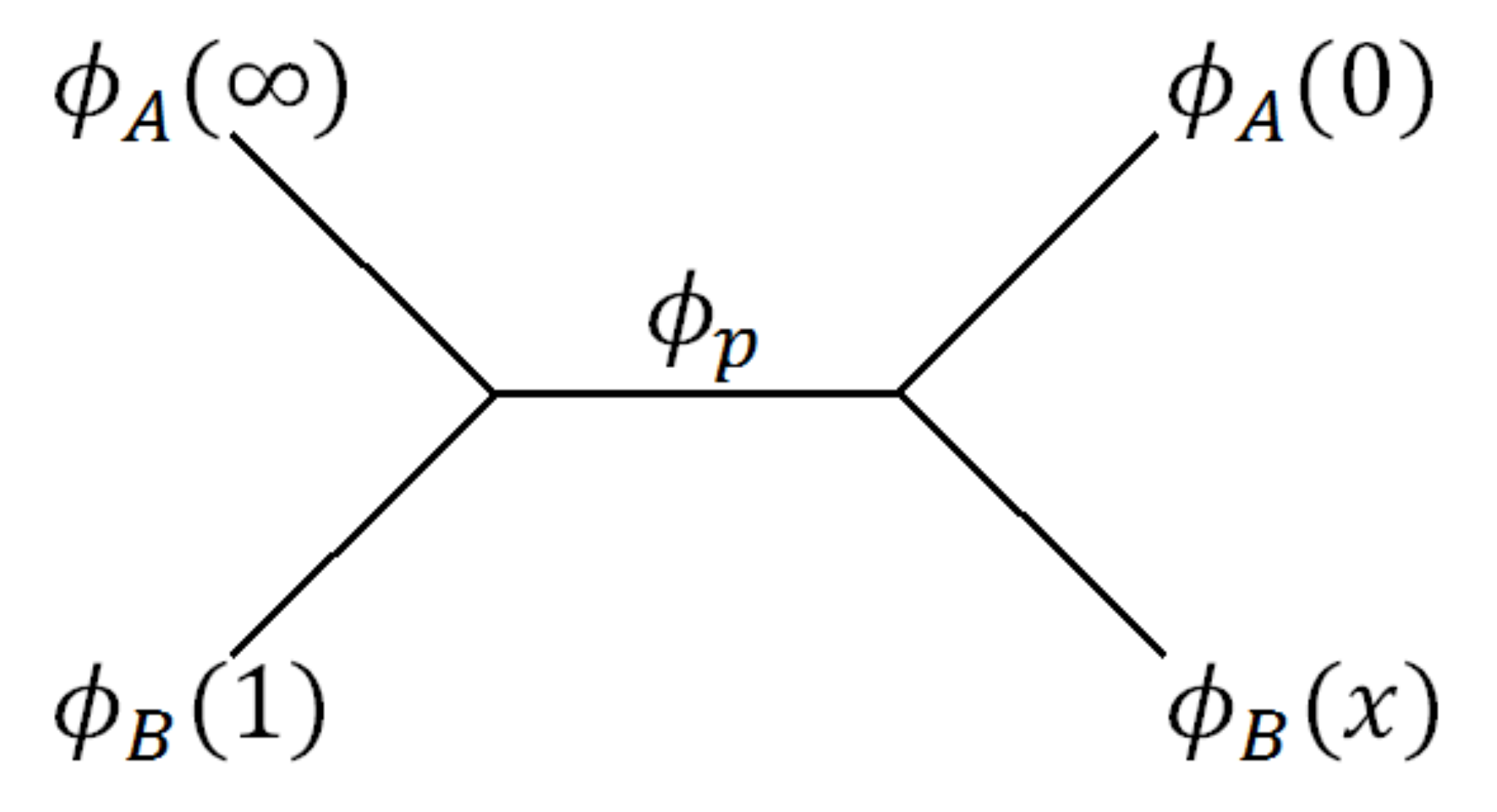}}
\newlength{\pdw}
\settowidth{\pdw}{\usebox{\boxpd}} 
\begin{equation*}
 \ca{F}^{BA}_{BA}(h_p|x) \equiv \parbox{\pdw}{\usebox{\boxpd}},
\end{equation*}
and as a result, we find 
\begin{enumerate}
\item heavy-light region: $h_A<h_B$ and $h_A<\fr{c}{32}$
\begin{equation}\label{eq:pre1}
\log \ca{F}^{BA}_{BA}(h_p|x)\ar{x \to 1} 
\pa{4h_A-2h_B-\fr{c-1}{6}\pa{1-\s{1-\fr{24}{c-1}h_A}}}\log(1-x) +O(\log\log(1-x)).
\end{equation}
\item heavy-light region: $h_B<h_A$ and $h_B<\fr{c}{32}$
\begin{equation}\label{eq:pre2}
\log \ca{F}^{BA}_{BA}(h_p|x)\ar{x \to 1} 
\pa{2h_B-\fr{c-1}{6}\pa{1-\s{1-\fr{24}{c-1}h_B}}}\log(1-x) +O(\log\log(1-x)).
\end{equation}
\item heavy-heavy region: $h_A, h_B>\fr{c}{32}$ 
\begin{equation}\label{eq:pre3}
\log \ca{F}^{BA}_{BA}(h_p|x)\ar{x \to 1}
\pa{\fr{c-1}{24}-2h_B}\log \pa{1-x}+   O(\log\log(1-x)).
\end{equation}
\end{enumerate}

The key point here is that a transition of the blocks along the lines $h_A, h_B=\fr{c}{32}$ was unknown before our recent studies \cite{Kusuki2018, Kusuki2018b}. However, these studies rely only on numerical calculations, therefore, there is no framework for further understanding of how this transition occurs.
The aim of this paper is to analytically derive (\ref{eq:pre1}) $\sim$ (\ref{eq:pre3}) (and also (\ref{eq:pre4}) and (\ref{eq:pre5})). In particular, our objective is to analytically understand the mechanism of the transition at $\fr{c}{32}$. The key to achieving this is to use the Zamolodchikov monodromy method \cite{Harlow2011,Zamolodchikov1987}, which is one of methods used to derive the large $c$ conformal blocks with heavy intermediate states. The original one gives only the blocks with very heavy intermediate states ($h_p \gg c$), but many improvements that are currently derived from it are introduced and can be used to estimate the blocks within various limits. In this paper, we study this monodromy method in the limit $x \to 1$, which is similar to the method used in \cite{Fitzpatrick2017}\footnote{We have to mention that they derive the equation (\ref{eq:trM}) for the special blocks, HHLL blocks, but in this paper, we give the equation for the most general blocks. Below (\ref{eq:trM}), they consider the late Lorentzian time limit, which is given by encircling the singular point infinite times. This limit is different from the limit $z \to 1$ without picking up a monodromy, which we are considering in this paper. As a result, we find the transition that we want.
}.
As a result, we perfectly reproduce our previous numerical results (\ref{eq:pre1}) $\sim$ (\ref{eq:pre3}) (and also (\ref{eq:pre4}) and (\ref{eq:pre5})) in the large $c$ limit. Particularly, we analytically confirm the existence of the transition at $\fr{c}{32}$. Our numerical works \cite{Kusuki2018, Kusuki2018b} and this analytic work robustly suggest that the large $c$ conformal blocks undergo a transition.
In the bulk, we can interpret it as a remarkable transition of the collision behavior between two particles.

In \cite{Kusuki2018, Kusuki2018b}, we conjecture that the asymptotic solutions to the Zamolodchikov recursion relation can be given by a simple expression in the form of a Cardy-like formula. This work can be thought of as one of analytic proof of this conjecture.

The outline of this paper is as follows. In Section \ref{sec:standard}, we begin with the monodromy method introduced by Zamolodchikov and derive the monodromy equation, the solutions of which give the conformal blocks. In section \ref{sec:mono}, we solve the monodromy equation in the vicinity of the singular point $x=1$ and, thereby, we analytically give the asymptotic expression of the conformal blocks conjectured in our recent work and also reveal the mechanism of the transition at $h_{A,B}=\fr{c}{32}$ for ABBA blocks in Section \ref{sec:block}. In Section \ref{sec:rec}, we explain how our results lead to the solutions to the Zamolodchikov recursion relation. In Section \ref{sec:heavy}, we study the conformal blocks with heavy intermediate states by using the same method and obtain the universal formula for the heavy-light-light OPE coefficients in large $c$ CFTs. We conclude with a discussion of our results in Section \ref{sec:discussion}.
\section{Standard Monodromy Method}\label{sec:standard}
In this section, we will explain the original monodromy method introduced by Zamolodchikov \cite{Zamolodchikov1987}. 
More details about this method can be found in the review \cite{Harlow2011,Suchanek2009}.

In this section, we will use the notation,
\newsavebox{\boxpe}
\sbox{\boxpe}{\includegraphics[width=160pt]{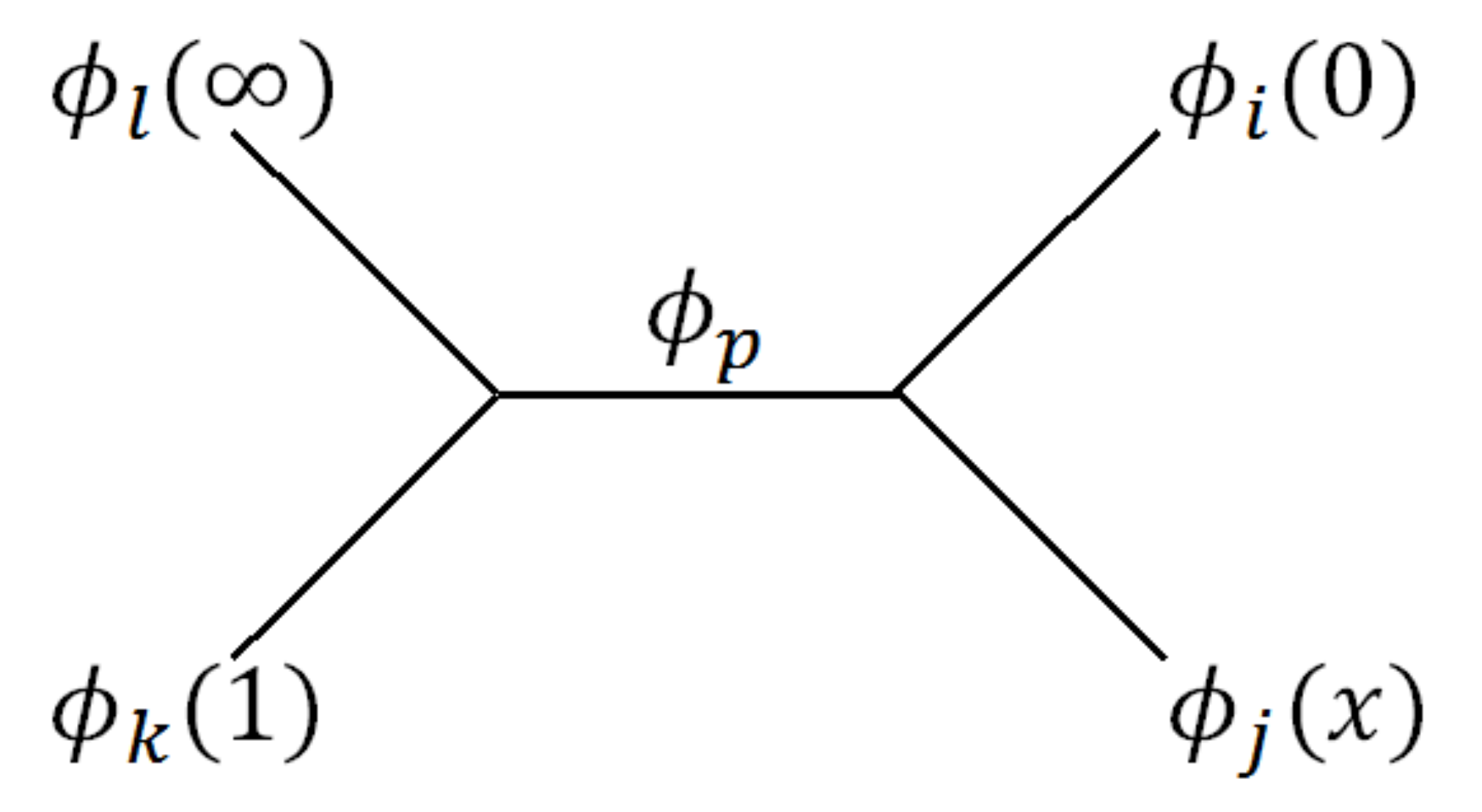}}
\newlength{\pew}
\settowidth{\pew}{\usebox{\boxpe}} 

\begin{equation*}
 \ca{F}^{ji}_{kl}(h_p|x) \equiv \parbox{\pew}{\usebox{\boxpe}},
\end{equation*}
and we define $c=1+6\pa{b+\fr{1}{b}}^2$, where the large $c$ limit corresponds to $b \to 0$.

\begin{figure}[h]
 \begin{center}
  \includegraphics[width=100mm]{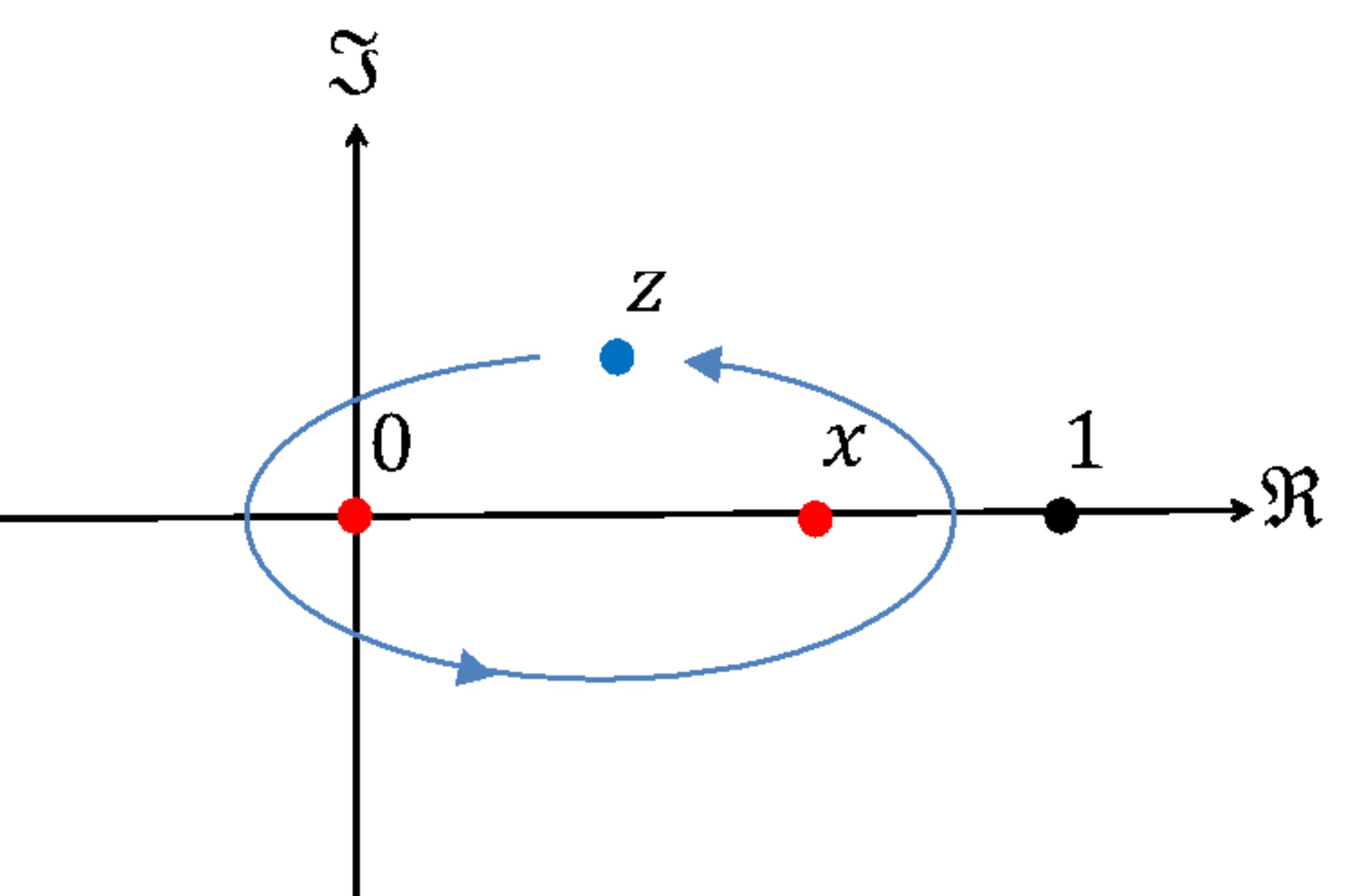}
 \end{center}
 \caption{ Since the operator $\ca{O}_p$ arises from the OPE between $\ca{O}_1$ and $\ca{O}_2$, the monodromy of $\Psi_p$ around $\ca{O}_p$ is given by encircling both points $z=\{0,x \}$ along a blue contour.}
 \label{fig:mono}
\end{figure}

\begin{enumerate}
\item {\it Null ordinary differential equation (ODE)}\\
	The degenerate primary operator $\Psi$ with the dimension $-\fr{1}{2}-\fr{3}{4}b^2$ leads to the ODE,
	\begin{equation}\label{eq:ODE}
	\left[\frac{1}{b^2}\partial_z^2+\sum_{i=1}^4\left(\frac{h_i}{(z-z_i)^2}+\frac{1}{z-z_i}\partial_i\right)\right]
	\langle\mathcal{O}_4(z_4,\bar{z}_4)\mathcal{O}_3(z_3,\bar{z}_3)\Psi(z,\bar{z})\mathcal{O}_2(z_2,\bar{z}_2)\mathcal{O}_1(z_1,\bar{z}_1)\rangle=0,
	\end{equation}
where we set $(z_1,z_2,z_3,z_4) \to (0,x,1,\infty)$.

\item {\it ODE for each intermediate state}\\
	Under some appropriate assumptions, which are reasonable in large $c$ CFTs, the ODE (\ref{eq:ODE}) leads to one ODE for each intermediate state $\ca{O}_p$ in the OPE $\ca{O}_1 \ca{O}_2$ as
	\begin{equation}\label{eq:ODE4}
	\left[\partial_z^2+\sum_{i=1}^4\left(\frac{\delta_i}{(z-z_i)^2}-\frac{C_i}{z-z_i}\right)\right]\Psi_p=0,
	\end{equation}
	where $\delta_i=b^2h_i$ and
	\begin{equation}
	\langle\mathcal{O}_4\mathcal{O}_3\Psi\mathcal{O}_p\rangle \equiv 
	\Psi_p(z,\bar{z};z_i,\bar{z}_i)\langle\mathcal{O}_4\mathcal{O}_3\mathcal{O}_p\rangle.
	\end{equation}
	At this stage, we cannot determine $C_i$, which is called {\it accessory parameter}. This parameter is related to the conformal block as
\begin{equation}
C_2=\partial_x f_{cl},
\end{equation}
where we assume that the large $c$ conformal blocks have the following form:
\begin{equation}\label{eq:semif}
\mathcal{F}^{21}_{34}(h_p|x)\sim \ex{-\frac{c}{6}f_{cl}},
\end{equation}
which is supported by the Liouville CFT \cite{Fitzpatrick2014}.

\item {\it Ward--Takahashi identity}\\
	The second term of (\ref{eq:ODE4}) can be understood as $b^2$ multiplied by the semiclassical expectation value of the stress tensor from the Ward--Takahashi identity. This leads to the following ODE:
	\begin{equation}\label{eq:ODE5}
	\left[\partial_z^2+\frac{\delta_1}{z^2}+\frac{\delta_2}{(z-x)^2}+\frac{\delta_3}{(1-z)^2}+\frac{\delta_1+\delta_2+\delta_3-\delta_4}{z(1-z)}-\frac{C_2 x(1-x)}{z(z-x)(1-z)}\right]\Psi_p=0.
	\end{equation}

\item {\it WKB approximation}\\
By using the WKB approximation in the limit $\d_p \to \infty$, we can solve the ODE (\ref{eq:ODE5}),
\begin{equation}\label{eq:Msolution}
\Psi_p\sim \exp\left[\pm\sqrt{x(1-x)C_2}\int_{z_0}^z\frac{dz'}{\sqrt{z'(1-z')(z'-x)}}\right].
\end{equation}

\item {\it Monodromy equation} \label{item:mono} \\
From the usual CFT discussion for degenerate operators, we obtain the OPE between $\ca{O}_p$ and $\Psi$ as
\begin{equation}
\pa{z-z_1}^{\fr{1}{2}\pa{1\pm\s{1-4b^2 h_p}}},
\end{equation}
therefore, the monodromy matrix of  $\Psi_p$  as $z$ circles both $x$ and $z_1$  (see Figure \ref{fig:mono})  in this basis is straightforwardly given by
\begin{equation}
\begin{aligned}
 M= \left(
    \begin{array}{cc}
    \ex{i \pi \pa{1 +\s{1-4b^2 h_p}} }   &  0 \\
     0  &  \ex{i \pi \pa{1 -\s{1-4b^2 h_p}} }   \\
    \end{array}
  \right).
\end{aligned}
\end{equation}

The solution (\ref{eq:Msolution}) needs to have the above monodromy, which leads to the following monodromy equation:
\begin{equation}\label{eq:C2}
C_2\simeq -\frac{\pi^2 b^2 h_p}{x(1-x)K(x)^2}.
\end{equation}

\item {\it Semiclassical conformal block}\\
Using the relation
\begin{equation}
C_2=\partial_x f_{cl},
\end{equation}
we can obtain the conformal blocks as
\begin{equation}
\mathcal{F}^{21}_{34}(h_p|x)=\pa{16q}^{h_p},\ \ \ \ \ \ q(x)=\ex{-\pi \fr{K(1-x)}{K(x)}}.
\end{equation}
Including the higher order leads to the well-known semi-classical block,
\begin{equation}\label{eq:semiclassical}
\mathcal{F}^{21}_{34}(h_p|x)=(16q)^{h_p-\frac{c-1}{24}}x^{\frac{c-1}{24}-h_1-h_2}(1-x)^{\frac{c-1}{24}-h_2-h_3}
(\theta_3(q))^{\frac{c-1}{2}-4(h_1+h_2+h_3+h_4)}.
\end{equation}
The technical detail of this derivation is shown in Appendix A of \cite{Suchanek2009}.
This method is called the {\it monodromy method}.
\end{enumerate}

Note that this method relies on the WKB method and, therefore, the range in which this method is valid is limited to
\begin{equation}\label{eq:WKBvalid}
h_p \abs{\log q}^2\gg c,
\end{equation}
which is explained in more details in \cite{Kusuki2018}.

\section{Monodromy Method at $x=1$}\label{sec:mono}

In Section \ref{sec:standard}, we used the solution approximated by the WKB method to obtain the monodromy equation. In this section, instead of using the WKB method, we take the limit of $x$ in (\ref{eq:ODE5}) to reduce it to the solvable ODE. This is a reasonable approach because what we want to do is to understand the asymptotic behavior of the blocks in the vicinity of $x=1$ as in(\ref{eq:pre1}) $\sim$ (\ref{eq:pre3}). 

We will start with the ODE (\ref{eq:ODE5}),
	\begin{equation}\label{eq:monoeq}
	\left[\partial_z^2+\frac{\delta_1}{z^2}+\frac{\delta_2}{(z-x)^2}+\frac{\delta_3}{(1-z)^2}+\frac{\delta_1+\delta_2+\delta_3-\delta_4}{z(1-z)}-\frac{\kappa}{z(z-x)(1-z)}\right]\Psi_p=0,
	\end{equation}
where we define $\kappa$ as $\kappa=C_2 x(1-x)$. Now that we want to estimate the asymptotics of the block at $x=1$, we take the limit $x \to 1$ in the ODE,
	\begin{equation}\label{eq:ODEatz=1}
	\left[z(1-z)\partial_z^2+\frac{\delta_1}{z}+\frac{\delta_2+\delta_3+\kappa}{1-z}-\d_4 \right]\Psi_p=0.
	\end{equation}
We define the parameters {$\eta_i$} as
\begin{equation}
\begin{aligned}
\left\{
    \begin{array}{ll}
       \d_1=\eta_1(1-\eta_1) ,\\
       \d_4=\eta_4(1-\eta_4) ,\\
       \d_2+\d_3+\kappa=\eta_{2+3+\kappa}(1-\eta_{2+3+\kappa}),\\
    \end{array}
  \right.\\
\end{aligned}
\end{equation}
which we call {\it effective Liouville momenta} because they are normalized Liouville momenta $\eta=b \a$ with $h=\a(Q-\a)$ in the large $c$ limit. Using this notation, we obtain the twisted hypergeometric equation
\begin{equation}
\br{z(1-z)\del_z^2+\fr{\eta_1(1-\eta_1)}{z}+\fr{\eta_{2+3+\kappa}(1-\eta_{2+3+\kappa})}{1-z} -\eta_4(1-\eta_4)}\Psi_p=0.
\end{equation}
The solutions to this equation are obtained by using the hypergeometric functions as
\begin{equation}\label{eq:sol1}
\begin{aligned}
\Psi_p^{(11)}&(z)=z^{\fr{C}{2}}(1-z)^{\fr{A+B-C+1}{2}} {}_2 F_1(A,B,C;z),\\
\Psi_p^{(21)}&(z)=z^{1-\fr{C}{2}}(1-z)^{\fr{A+B-C+1}{2}} {}_2 F_1(A-C+1,B-C+1,2-C;z),
\end{aligned}
\end{equation}
where $A,B$ and $C$ are related to the effective Liouville momenta as follows:
\begin{equation}
\begin{aligned}
\left\{
    \begin{array}{ll}
      &A=-1+\eta_1+\eta_4+\eta_{2+3+\kappa}, \\
      &B=\eta_1-\eta_4+\eta_{2+3+\kappa}, \\
      &C=2 \eta_1. \\
    \end{array}
  \right.\\
\end{aligned}
\end{equation}

\begin{figure}[h]
 \begin{center}
  \includegraphics[width=120mm]{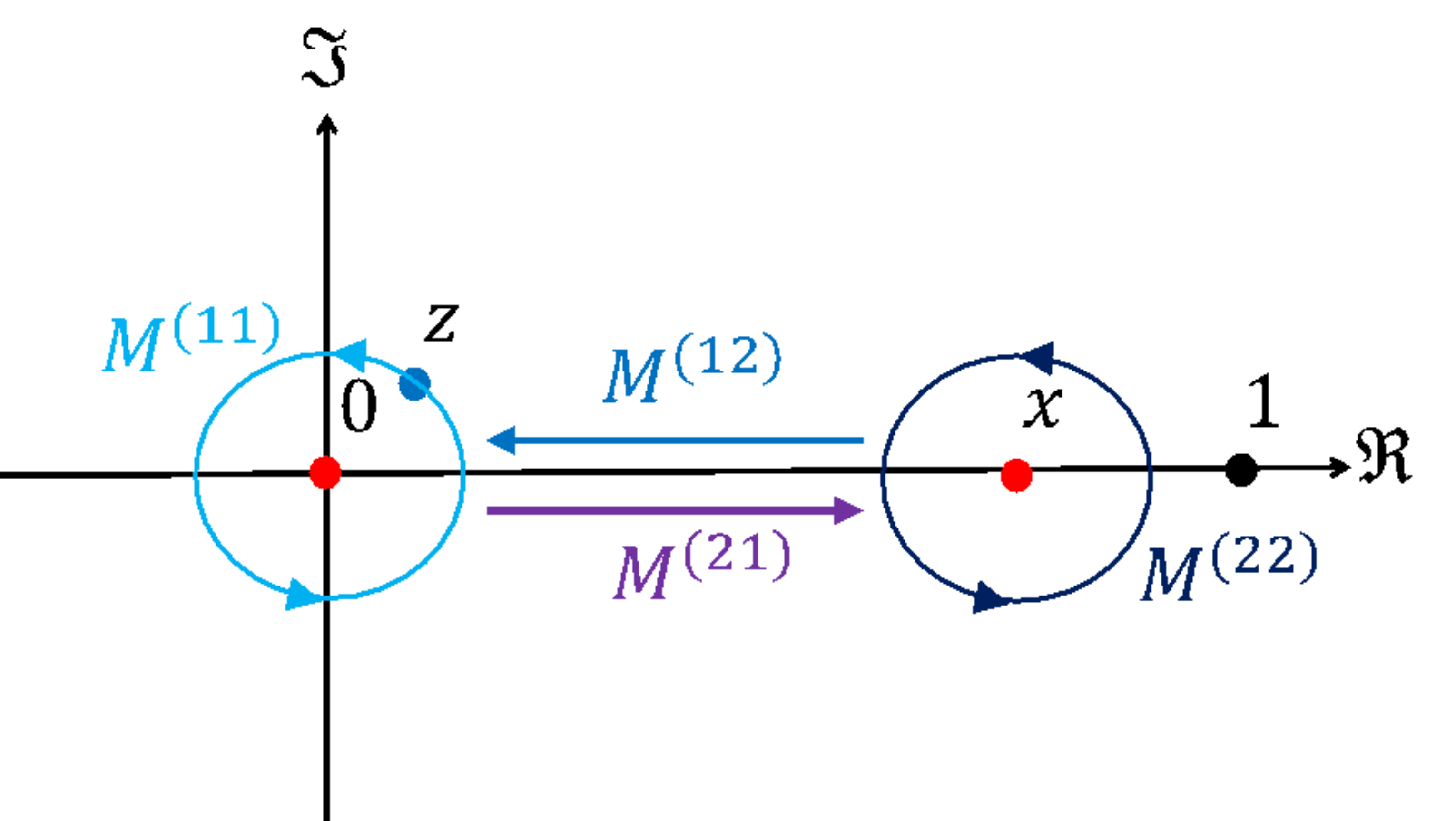}
 \end{center}
 \caption{To obtain the monodromy matrix, we have to evaluate the monodromy around $z=x$, while excluding $z=1$. However, as long as we are restricted to the solution of the ODE (\ref{eq:ODEatz=1}), we cannot distinguish the points $z=\{x,1\}$. Therefore, we have to see the solution to the ODE in another limit $1-x \ll 1$ with $\fr{z-x}{1-x} \gg 1$, which is shown in (\ref{eq:monoeq2}). The matrices displayed in this figure correspond to the matrices (\ref{eq:M12}), (\ref{eq:M11}), and the whole circle corresponds to (\ref{eq:MMMM}).  }
 \label{fig:mono2}
\end{figure}

To obtain the monodromy matrix, we have to know the solutions to the OPE in another limit $1-x \ll 1$ with $\fr{z-x}{1-x}$ fixed (see Figure \ref{fig:mono2}). In this region, it is useful to change variables in the ODE (\ref{eq:monoeq}) to $w=\fr{z-x}{1-x}$. In the limit $x \to 1$, the ODE reduces to
	\begin{equation}\label{eq:monoeq2}
	\left[w(1-w)\partial_w^2+\frac{\delta_2}{w}+\frac{\delta_3}{1-w}-(\kappa+\d_2+\d_3) \right]\Psi_p=0.
	\end{equation}
Using the notation,
\begin{equation}
\begin{aligned}
\left\{
    \begin{array}{ll}
       \d_2=\eta_2(1-\eta_2) ,\\
       \d_3=\eta_3(1-\eta_3) ,\\
       \d_2+\d_3+\kappa=\eta_{2+3+\kappa}(1-\eta_{2+3+\kappa}),\\
    \end{array}
  \right.
\left\{
    \begin{array}{ll}
      &A'=-1+\eta_2+\eta_3+\eta_{2+3+\kappa}, \\
      &B'=\eta_2+\eta_3-\eta_{2+3+\kappa}, \\
      &C'=2 \eta_2, \\
    \end{array}
  \right.\\
\end{aligned}
\end{equation}
The solutions to the ODE (\ref{eq:monoeq2}) are also given by the hypergeometric functions as: 
\begin{equation}\label{eq:sol2}
\begin{aligned}
\Psi_p^{(12)}&(w)=w^{\fr{C'}{2}}(1-w)^{\fr{A'+B'-C'+1}{2}} {}_2 F_1(A',B',C';w),\\
\Psi_p^{(22)}&(w)=w^{1-\fr{C'}{2}}(1-w)^{\fr{A'+B'-C'+1}{2}} {}_2 F_1(A'-C'+1,B'-C'+1,2-C';w).
\end{aligned}
\end{equation}

Both of the solutions $\pa{\Psi_p^{(11)}(z),  \Psi_p^{(21)}(z) }$ and $\pa{\Psi_p^{(12)}(w),  \Psi_p^{(22)}(w) }$ are valid in the special region $1-z \ll 1$ with $\fr{z-x}{1-x} \gg 1$. Therefore, we can find a way to change the basis from $\pa{\Psi_p^{(12)}(w),  \Psi_p^{(22)}(w) }$ to $\pa{\Psi_p^{(11)}(z),  \Psi_p^{(21)}(z) }$ by matching the two bases in this region. Thus, we can find the basis change matrix explicitly and express it as
\begin{equation}\label{eq:M12}
\begin{aligned}
  \left(
    \begin{array}{c}
       \Psi_p^{(11)}(z)   \\
       \Psi_p^{(21)}(z)   \\
    \end{array}
  \right)
=M^{(12)}
  \left(
    \begin{array}{c}
       \Psi_p^{(12)}(w)   \\
       \Psi_p^{(22)}(w)   \\
    \end{array}
  \right).
\end{aligned}
\end{equation}
From the expressions (\ref{eq:sol1}) and (\ref{eq:sol2}), the monodromy matrices around $z=0$ and $w=0$ are simply given by
\begin{equation}\label{eq:M11}
\begin{aligned}
M^{(11)} =
   \left(
    \begin{array}{cc}
    \ex{\pi i C}   & 0  \\
     0  &  \ex{-\pi i C}  \\
    \end{array}
  \right),
\ \ \ \ \ \ \ \ 
M^{(22)} =
   \left(
    \begin{array}{cc}
    \ex{\pi i C'}   & 0  \\
     0  &  \ex{-\pi i C'}  \\
    \end{array}
  \right).
\end{aligned}
\end{equation}
At this stage, the monodromy matrix encircling both $z=\{0, x\}$ is given by
\begin{equation}\label{eq:MMMM}
M=M^{(11)}M^{(12)}M^{(22)}M^{(21)},
\end{equation}
where $M^{(21)}$ is the inverse matrix of $M^{(12)}$. The circle corresponding to each matrix is sketched in Figure \ref{fig:mono2}.  The matrix $M$ is very complicated but all we need to know is only its trace because the determinant of $M$ is trivially $\det M=1$. The trace of $M$ can be expressed as
\begin{equation}\label{eq:trM}
\begin{aligned}
\tr M
=&-\ca{M}(A,B,C) \ca{M}(A',C'-B',C') \e^{2 \ek-1}\\
&-\ca{M}(C-A,C-B,C) \ca{M}(C'-A',B',C') \e^{-(2 \ek-1)}+\ca{M}_{const},
\end{aligned}
\end{equation}
where $\epsilon \equiv 1-x$ and $\ca{M}$ and $\ca{M}_{const}$ are constants in $\e$ and defined as follows:
\begin{equation}
\begin{aligned}
\ca{M}(A,B,C)&=2\pi \fr{\G(C-A-B)\G(C-A-B+1)}{\G(1-A)\G(1-B)\G(C-A)\G(C-B)},\\
\ca{M}_{const} &=2+4\Biggl( \fr{\sin\pa{\pi(C-A)}\sin(\pi(C-B))\sin(\pi A')\sin(\pi(C'-B'))}{\sin(\pi(A+B-C))\sin(\pi(A'-B'))}\\
& \hspace{7ex} +\fr{\sin\pa{\pi A}\sin(\pi B)\sin(\pi (C'-A')\sin(\pi B')}{\sin(\pi(A+B-C))\sin(\pi(A'-B'))} \Biggr).
\end{aligned}
\end{equation}
Note that this function $\ca{M}$ has the following properties:
\begin{equation}
\begin{aligned}
\ca{M}(A,B,C)&=\ca{M}(B,A,C),\\
\ca{M}(C-A,C-B,C)&=\ca{M}(1-A,1-B,2-C),\\
\end{aligned}
\end{equation}
The monodromy equation (corresponding to \ref{item:mono}. in Section \ref{sec:standard}) is
\begin{equation}\label{eq:trM=}
\tr M= -2 \cos(\pi \a_p),
\end{equation}
where $\a_p=\s{1-\fr{24}{c}h_p}$. It can be observed that, in the limit $\e \to 0$, either $\e^{2 \ek-1}$ or $\e^{-(2 \ek-1)}$ diverges in the left hand side of (\ref{eq:trM}), while the right hand side is finite. This implies that the parameters $\{A,B,C\}$ (in other words, the accessory parameter $\kappa$) are constrained by one of the following three ways:
\begin{description}
\item[(i)] $\ca{M}(A,B,C) \ca{M}(A',C'-B',C') =0$ and $\abs{\e^{-(2 \ek-1)}} \ar{\e \to 0} 0$.
\item[(i')] $\ca{M}(C-A,C-B,C) \ca{M}(C'-A',B',C')=0$ and $\abs{\e^{2 \ek-1}}\ar{\e \to 0}  0$.
\item[(ii)] $\Re(2\ek-1)=0$.
\end{description}

We will start with the case (i). The condition$\abs{\e^{-(2 \ek-1)}} \ar{\e \to 0} 0$ means that
\begin{equation}\label{eq:ek}
\Re\ek<\fr{1}{2}.
\end{equation}
In addition, the equation $\ca{M}(A,B,C) \ca{M}(A',C'-B',C') =0$ is realized by the singularities of the Gamma functions in the denominator of $\ca{M}$ (by the property $1/\G(-n)=0$ for $n \in \bb{Z}_{\geq 0}$). As a result, the condition (i) leads to
\begin{equation}\label{eq:eksol}
\begin{aligned}
\ek&=\left\{
    \begin{array}{ll}
     -(\eta_1+\eta_4)+2+\bb{Z}_{\geq 0}  ,& \\
     \pm(\eta_1-\eta_4)+1+\bb{Z}_{\geq 0}  ,&   \\
     \eta_1+\eta_4+\bb{Z}_{\geq 0}  ,& \\
     -(\eta_2+\eta_3)+2+\bb{Z}_{\geq 0}  ,& \\
     \pm(\eta_2-\eta_3)+1+\bb{Z}_{\geq 0}  ,&   \\
     \eta_2+\eta_3+\bb{Z}_{\geq 0}  ,& \\
    \end{array}
  \right.\\
\end{aligned}
\end{equation}
where $\bb{Z}_{\geq 0}$ indicates the positive integer set $\{0,1,2,...\} $. Since the Liouville momenta $\eta_1, \eta_2, \eta_3, \eta_4$ satisfy the inequality $0 \leq \Re \eta_i \leq \fr{1}{2}$ in unitary CFTs with  (\ref{eq:ek}) known, one can deduce that only the two solutions in (\ref{eq:eksol}) are allowed. The two solutions are 
\begin{equation}\label{eq:kappasol}
\begin{aligned}
\ek&=\left\{
    \begin{array}{ll}
     \eta_1+\eta_4  ,& \\
     \eta_2+\eta_3  ,& \\
    \end{array}
  \right.\\
\end{aligned}
\end{equation}
which means that the accessory parameter $\kappa$ is given by
\begin{equation}
\begin{aligned}
\kappa&=\left\{
    \begin{array}{ll}
      -\d_2-\d_3+\d_1+\d_4-2\eta_1 \eta_4  ,& \\
     -2\eta_2 \eta_3  .& \\
    \end{array}
  \right.\\
\end{aligned}
\end{equation}
Note that these solutions exist only if the inequality (\ref{eq:ek}) is satisfied. Combining this inequality with (\ref{eq:kappasol}) leads to 
\begin{equation}\label{eq:kappa}
\begin{aligned}
\kappa&=\left\{
    \begin{array}{ll}
      -\d_2-\d_3+\d_1+\d_4-2\eta_1 \eta_4  ,& \text{if} \ \ \ \eta_1+\eta_4<\fr{1}{2}, \\
     -2\eta_2 \eta_3  ,& \text{if} \ \ \ \eta_2+\eta_3<\fr{1}{2}. \\
    \end{array}
  \right.\\
\end{aligned}
\end{equation}
This is one of the important results presented in this paper.

The case (i') is almost equivalent to the case (i).  The conditions $\abs{\e^{2 \ek-1}}\ar{\e \to 0}  0$ and $\ca{M}(C-A,C-B,C) \ca{M}(C'-A',B',C')=0$ lead to
\begin{equation}\label{eq:ek2}
\Re\ek>\fr{1}{2},
\end{equation}
and
\begin{equation}\label{eq:eksol2}
\begin{aligned}
\ek&=\left\{
    \begin{array}{ll}
     -(\eta_1+\eta_4)+1-\bb{Z}_{\geq 0}  ,& \\
     \pm(\eta_1-\eta_4)-\bb{Z}_{\geq 0}  ,&   \\
     \eta_1+\eta_4-1-\bb{Z}_{\geq 0}  ,& \\
     -(\eta_2+\eta_3)+1-\bb{Z}_{\geq 0}  ,& \\
     \pm(\eta_2-\eta_3)-\bb{Z}_{\geq 0}  ,&   \\
     \eta_2+\eta_3-1-\bb{Z}_{\geq 0}  .& \\
    \end{array}
  \right.\\
\end{aligned}
\end{equation}

The inequalities for $\ek$ and $\eta_{1,2,3,4}$ restrict the solutions to
\begin{equation}
\begin{aligned}
\ek&=\left\{
    \begin{array}{ll}
     -(\eta_1+\eta_4)+1 ,& \\
     -(\eta_2+\eta_3)+1  ,& \\
    \end{array}
  \right.\\
\end{aligned}
\end{equation}
and, consequently, we obtain the accessory parameter as
\begin{equation}
\begin{aligned}
\kappa&=\left\{
    \begin{array}{ll}
      -\d_2-\d_3+\d_1+\d_4-2\eta_1 \eta_4  ,& \text{if} \ \ \ \eta_1+\eta_4<\fr{1}{2}, \\
     -2\eta_2 \eta_3  ,& \text{if} \ \ \ \eta_2+\eta_3<\fr{1}{2}, \\
    \end{array}
  \right.\\
\end{aligned}
\end{equation}
which is exactly the same as (\ref{eq:kappa}).

In the case (ii), one finds that $\ca{M}$ and $\ca{M}_{const}$ diverge and, therefore, putting the condition (ii) into (\ref{eq:trM})  seems to be ill-defined.
However, one can show that 
\begin{equation}
\tr M \ar{2\ek-1=\w\to0} O(\w^0),
\end{equation}
which is well-defined. Although we can explicitly show the constant part of the limit of the trace, we do not present it here as it is beyond the scope of this paper. 

Since the trace of the monodromy matrix should be a real constant, if we express the condition (ii) as
\begin{equation}
\ek=\fr{1}{2}+iP,
\end{equation}
then $P$ can take only the values 
\begin{equation}
P=\fr{\pi m}{2 \log \e} \ar{\e \to 0} 0, \ \ \ \ \ \ \ m\in \bb{Z}.
\end{equation}
As a result, we obtain the accessory parameter as
\begin{equation}
\kappa=-\d_2-\d_3+\fr{1}{4}.
\end{equation}
\section{Conformal Blocks at $x=1$}\label{sec:block}

In Section \ref{sec:mono}, we obtain the accessory parameter for large $c$ conformal blocks at $x=1$. Before moving on to the next step, we summarize the results as follows:
\begin{equation}\label{eq:sumk}
\begin{aligned}
\kappa&=\left\{
    \begin{array}{ll}
      -\d_2-\d_3+\d_1+\d_4-2\eta_1 \eta_4  ,& \text{if} \ \ \ \eta_1+\eta_4<\fr{1}{2}, \\
     -2\eta_2 \eta_3  ,& \text{if} \ \ \ \eta_2+\eta_3<\fr{1}{2}, \\
    -\d_2-\d_3+\fr{1}{4}. \\
    \end{array}
  \right.\\
\end{aligned}
\end{equation}
From this accessory parameter, we can reconstruct the conformal blocks by
\begin{equation}
\ca{F}^{21}_{34}(h_p|x=1-\e)\ar{\e \to 0} \e^{\fr{c}{6} \kappa}.
\end{equation}
However, which of the expressions in (\ref{eq:sumk}) should we use to obtain the conformal blocks?
All of them are solutions of the monodromy equation. Therefore, we expect that the conformal blocks are given by the summation of these accessory parameters. Nevertheless, in the limit $c \log \fr{1}{\e} \gg 1$ (large $c$ limit or $x \to 1$ limit), the leading conformal blocks are given by the minimum of (\ref{eq:sumk}). For convenience, we label each accessory parameter as
\begin{equation}
\kappa_1=-\d_2-\d_3+\d_1+\d_4-2\eta_1 \eta_4 , \ \ \ \ \ \ 
\kappa_2 =  -2\eta_2 \eta_3 , \ \ \ \ \ \ 
\kappa_3=-\d_2-\d_3+\fr{1}{4}.
\end{equation}
We can show the following relations for these parameters:
\begin{equation}
\begin{aligned}
\kappa_3-\kappa_1&=\pa{\eta_1+\eta_4-\fr{1}{2}}^2>0,  \hspace{26ex}  \text{if} \ \ \ \eta_1+\eta_4<\fr{1}{2}, \\
\kappa_3-\kappa_2&=\pa{\eta_2+\eta_3-\fr{1}{2}}^2>0,    \hspace{26ex}  \text{if} \ \ \ \eta_2+\eta_3<\fr{1}{2}, \\
\kappa_1-\kappa_2&=(\eta_1+\eta_4-\eta_2-\eta_3)(1-\eta_1-\eta_2-\eta_3-\eta_4), \ \ \ \ \ \ \ \ \ \ \
   \text{if} \ \ \ \eta_1+\eta_4<\fr{1}{2}, \eta_2+\eta_3<\fr{1}{2}. \\
\end{aligned}
\end{equation}
This means that if the inequality $\eta_1+\eta_4<\fr{1}{2}$ (or $\eta_2+\eta_3<\fr{1}{2}$) is satisfied, then the leading conformal blocks are provided by $\kappa_1$ (or $\kappa_2$) and, otherwise, they are given by $\kappa_3$.

We will set $\eta_1=\eta_4=\eta_A$ and $\eta_2=\eta_3=\eta_B$, which correspond to the ABBA blocks. In this case, we can obtain the leading conformal blocks as:
\begin{equation}\label{eq:ABBA}
\begin{aligned}
\log \ca{F}^{BA}_{BA}(h_p|x=1-\e)\ar{\e \to 0} &\left\{
    \begin{array}{ll}
     \pa{ 4h_A-2h_B-\fr{c}{3}\eta_A }\log \e ,& \text{if} \ \ \ \eta_A<\fr{1}{4} \text{ and } \eta_A<\eta_B , \\
    \pa{2h_B-\fr{c}{3}\eta_B }\log \e  ,& \text{if} \ \ \ \eta_B<\fr{1}{4}  \text{ and } \eta_A>\eta_B, \\
   \pa{\fr{c}{24} -2h_B }\log \e  ,& \text{ otherwise},\\
    \end{array}
  \right.\\
\end{aligned}
\end{equation}
where $\eta_i=\fr{1-\s{1-\fr{24}{c}h_i}}{2}$. The inequality $\eta_i <\fr{1}{4}$ can be written in terms of the conformal dimensions as
\begin{equation}
h_i<\fr{c}{32},
\end{equation}
which is exactly the value we seek. Moreover, the results (\ref{eq:ABBA}) perfectly reproduce our recent results (\ref{eq:pre1}) $\sim$ (\ref{eq:pre3}) in the large $c$ limit. Thus, we can analytically show the transition of ABBA blocks at $h_{A,B}=\fr{c}{32}$.

In the case where $\eta_1=\eta_4=\eta_A$ and $\eta_2=\eta_3=\eta_B$, the large $c$ conformal blocks can be expressed by
\begin{equation}\label{eq:AABB}
\begin{aligned}
\log \ca{F}^{AA}_{BB}(h_p|x=1-\e)\ar{\e \to 0} &\left\{
    \begin{array}{ll}
     \pa{ -\fr{c}{3}\eta_A \eta_B }\log \e ,& \text{if} \ \ \ \eta_A+\eta_B<\fr{1}{2} , \\
   \pa{\fr{c}{24} -h_A-h_B }\log \e  ,& \text{ otherwise},\\
    \end{array}
  \right.\\
\end{aligned}
\end{equation}
which is also consistent with our numerical results (\ref{eq:pre4}) $\sim$  (\ref{eq:pre5}).
\section{General Solutions to the Recursion Relation}\label{sec:rec}
We can argue that the result in Section \ref{sec:block} is the analytic proof of our conjecture for the solution to the Zamolodchikov recursion relation \cite {Kusuki2018,Kusuki2018b}. The Zamolodchikov recursion relation is one of the tools used to calculate the conformal blocks numerically and has been recently receiving much attention \cite{Chen2017,Ruggiero2018} because it effectively encompasses the conformal blocks beyond the known regimes or limits.
\footnote{The Zamolodchikov recursion relation is also used in the conformal bootstrap \cite{EsterlisFitzpatrickRamirez2016,BaeLeeLee2016, LinShaoSimmons-DuffinWangYin2017,CollierKravchukLinYin2017}. The reference \cite{Perlmutter2015} presents a good review of this matter, and discusses the connections between various recursion relations.
 A generalization of the recursion relation to more general Riemann surfaces is given in \cite{Cho2017a}.
}
Next, we will very briefly explain the recursion relation and our recent conjecture in \cite {Kusuki2018,Kusuki2018b}. We will also relate this conjecture to the large $c$ conformal blocks derived in this paper.

By decomposing the conformal blocks into two parts, we obtain:
\begin{equation}
\ca{F}^{21}_{34}(h_p|x)=\Lambda^{21}_{34}(h_p|q)H^{21}_{34}(h_p|q),\ \ \ \ \ \ q(x)=\ex{-\pi \fr{K(1-x)}{K(x)}},
\end{equation}
where the function $\Lambda^{21}_{34}(h_p|q)$ is a universal prefactor given by
\begin{equation}\label{eq:pre}
 \Lambda^{21}_{34}(h_p|q)=(16q)^{h_p-\frac{c-1}{24}}x^{\frac{c-1}{24}-h_1-h_2}(1-x)^{\frac{c-1}{24}-h_2-h_3}
(\theta_3(q))^{\frac{c-1}{2}-4(h_1+h_2+h_3+h_4)}.
\end{equation}
The function $H^{21}_{34}(h_p|q)$ can be calculated recursively by using the following relation:
\begin{equation}
H^{21}_{34}(h_p|q)=1+\sum^\infty_{m=1,n=1}\frac{q^{mn}R_{m,n}}{h_p-h_{m,n}}H^{21}_{34}(h_{m,n}+mn|q),
\end{equation}
where $R_{m,n}$ is a constant in $q$, which is defined by
\begin{equation}\label{eq:Rmn}
R_{m,n}=2\fr{
\substack{m-1\\ \displaystyle{\prod} \\p=-m+1\\p+m=1 (\text{mod } 2) \  } \ 
\substack{n-1\\ \displaystyle{\prod} \\q=-n+1\\q+n=1 (\text{mod } 2) }
\pa{\la_2+\la_1-\la_{p,q}}\pa{\la_2-\la_1-\la_{p,q}}\pa{\la_3+\la_4-\la_{p,q}}\pa{\la_3-\la_4-\la_{p,q}}}
{\substack{
\substack{m \\ \displaystyle{\prod} \\k=-m+1 } \ \ 
\substack{n \\ \displaystyle{\prod} \\l=-n+1 }\\
(k,l)\neq(0,0), (m,n)
}
 \la_{k,l}}.
\end{equation}
In the above expressions, we used the notation
\begin{equation}
\begin{aligned}
&c=1+6\pa{b+\fr{1}{b}}^2,  \hspace{16ex}   h_i=\fr{c-1}{24}-\l_i^2,\\
&h_{m,n}=\fr{1}{4}\pa{b+\fr{1}{b}}^2-\lambda_{m,n}^2,  \hspace{10ex}  \lambda_{m,n}=\fr{1}{2} \pa{\fr{m}{b}+nb}.
\end{aligned}
\end{equation}

In our previous works \cite{Kusuki2018,Kusuki2018b}, we provided general solutions to this recursion relation by numerical computations.
If we re-express the function $H^{21}_{34}(h_p|q)$ as
\be
H^{21}_{34}(h_p|q)=1+\sum_{k=1}^\infty c_k(h_p) q^{k},
\ee
and the corresponding recursion relation as
\begin{equation}\label{eq:ck}
	c_k(h_p) = \sum_{i=1}^k \sum_{\substack{m=1, n=1\\mn=i}} \frac{R_{m,n}}{h_p - h_{m,n}} c_{k-i}(h_{m,n}+mn),
\end{equation}
then the solution $c_n$ for large $n$ takes the simple {\it Cardy-like} form of
	\begin{equation}\label{eq:cn}
		c_n \sim n^{\a} \ex{A \s{n}}.
	\end{equation}
We obtain $A$ and $\a$ for the ABBA blocks as follows:
\begin{enumerate}
\item heavy-heavy region: $h_A, h_B>\fr{c}{32}$ 
\begin{equation}\label{eq:AaHHABBA}
\begin{aligned}
A&=0,\\
\a&=4(h_A+h_B)-\fr{c+9}{4}.
\end{aligned}
\end{equation}

\item heavy-light region: $h_A>h_B$ and $h_B<\fr{c}{32}$
\begin{equation}\label{eq:AaHLABBA}
\begin{aligned}
A&=2\pi \sqrt{ \frac{c-1}{24}-4h_B+\fr{c-1}{6}\pa{1-\s{1-\fr{24}{c-1}h_B}}},\\
\a&=2(h_A+h_B)-\fr{c+5}{8}.
\end{aligned}
\end{equation}
\end{enumerate}
Note that, by the recursion relation, the coefficients $c_n$ for the ABBA blocks are symmetric when exchanging $h_A \leftrightarrow h_B$.

Essentially, these solutions lead perfectly to the ABBA blocks (\ref{eq:ABBA}) in the large $c$ limit. However, in \cite{Kusuki2018,Kusuki2018b} the expressions for ABBA blocks and their transition at $h_{A,B}=\fr{c}{32}$ came only from numerical observations. Therefore, their mechanism was not clear. In this paper, we are able to analytically determine $A$ and $\a$ in terms of the coefficients of the form (\ref{eq:cn}) and understand the mechanism of the transition by using the monodromy method. In fact, these asymptotic properties of the solutions to the recursion relation are highly nontrivial from the recursion relation itself. This is attributed to the fact that these asymptotic forms may come from numerous cancellations between many terms in the sum (\ref{eq:ck}). It is also important to understand these properties from the recursion relation itself. However, we leave that to future work.

\section{Conformal Blocks with Heavy Intermediate States}\label{sec:heavy}
When considering applications for conformal bootstrap, the region in which the above monodromy method can be applied must be taken into account. This issue will be discussed in this section, but relevant discussion can be found in \cite{Cardy2017}.

First, we have to precisely determine the regime of validity of the monodromy method discussed in Section \ref{sec:mono}. The discussion below (\ref{eq:trM=}) is valid only if the right hand side in (\ref{eq:trM=}) effectively vanishes, compared to the left hand side. More precisely, it is valid only if
\begin{equation}
\fr{h_p}{(\log \e)^2} \ll 1.
\end{equation}

What would happen if we go outside this region?
If the intermediate state is very heavy $h_p \gg c$, then the monodromy equation (\ref{eq:trM=}) reduces to
\begin{equation}
\e^{\abs{2\ek-1}} \sim \ex{-\pi i \a_p},
\end{equation}
where the symbol ``$\sim$'' means an approximation up to a constant factor. This equation leads to the accessory parameter
\begin{equation}
\kappa = \fr{1}{4}-\d_2-\d_3-\fr{\pi^2}{4(\log \e)^2}\pa{\fr{24}{c} h_p-1},
\end{equation}
when we restrict the analysis to the regime $ \fr{h_p}{(\log \e)^2} \nll 1$, where the term of order $O(\fr{1}{(\log \e)^2})$ remains even in the limit $\e \to 0$, unlike the method in Section \ref{sec:mono}. As a result, we obtain the conformal blocks with very heavy intermediate states as
\begin{equation}
\log \ca{F}^{21}_{34}(h_p|x=1-\e) = \pa{\fr{c}{24}-h_2-h_3} \log \e + \pi^2\pa{h_p-\fr{c}{24}}\fr{1}{\log \e}
\end{equation}
where we use $\int  \fr{\dd \e}{\e(\log \e)^2}=-\fr{1}{\log \e}$. This asymptotic expression of the conformal blocks exactly matches the semiclassical conformal blocks (\ref{eq:semiclassical}) in the limit $x \to 1$ since the elliptic nome has the following asymptotics,
\begin{equation}\label{eq:asell}
q(x)=\ex{-\pi \fr{K(1-x)}{K(x)}} \ar{1-x=\e\to0} \ex{\pi^2 \fr{1}{\log \e}},
\end{equation}
where we use $K(x)\ar{x \to 0}\fr{\pi}{2}$ and $K(x) \ar{x \to 1}-\fr{\log \pa{1-x}}{2}$.
This result implies that we can go beyond the initial regime (\ref{eq:WKBvalid}). It means that, in the limit $\e \to 1$ and $h_p \to \infty$ with $h_p/ (\log \e)^2 \nll 1$, we have
\begin{equation}\label{eq:monovalid}
H^{21}_{34}(h_p|q) \ar{h_p \to \infty} 1.
\end{equation}
In \cite{Kusuki2018} (and \cite{Das2017} in a special case), the conformal bootstrap equation for a four-point correlator in a sphere is discussed in the {\it high low temperature limit}, which is named as such due to the fact that if we re-express the elliptic nome $q(x)$ as $q=\ex{-\beta}$ as usual, then the limit $x\to 0$ vs. $x \to 1$ for the conformal boostrap equation can be regarded as the modular bootstrap equation in the high low temperature limit. The bootstrap equation re-expressed by the temperature is derived in \cite{Kusuki2018} as
\begin{equation}\label{eq:bootstrap}
\begin{aligned}
&\sum_{p}C_{AAp}C_{BBp}
\pa{\pa{16}^{h_p} \ex{-\fr{\beta}{2}\pa{h_p-\fr{c-1}{24}}} H^{AA}_{BB}(h_p|q) }
\pa{\pa{16}^{\bar{h}_p} \ex{-\fr{\bar{\beta}}{2}\pa{\bar{h}_p-\fr{c-1}{24}}} \overline{H^{AA}_{BB}}(\bar{h}_p|\bar{q}) }\\
&=
\sum_{p}C_{ABp}^2  
\pa{\pa{\fr{\beta}{2\pi}}^{\fr{c-1}{4}-4(h_A+h_B)}\pa{16}^{h_p}
\ex{-\fr{2\pi^2}{\beta}\pa{h_p-\fr{c-1}{24}}} 
H^{AB}_{AB}(h_p|\ti{q})}\\
& \hspace{10ex} \times\Biggl( \pa{\fr{\bar{\beta}}{2\pi}}^{\fr{c-1}{4}-4(\bar{h}_A+\bar{h}_B)}\pa{16}^{\bar{h}_p}
 \ex{-\fr{2\pi^2}{\bar{\beta}}\pa{\bar{h}_p-\fr{c-1}{24}}} 
\overline{H^{AB}_{AB}}(\bar{h}_p|\ti{\bar{q}}) \Biggr)  ,
\end{aligned}
\end{equation}
where $\ti{q}(x)=q(1-x)$. Here, we set $q=\ex{-\fr{\beta}{2}}$ and $\bar{q}=\ex{-\fr{\bar{\beta}}{2}}$, and we end up with $H^{21}_{34}(h_p|q) \ar{q \to 0} 1$ and  (\ref{eq:monovalid}). Hence, in the high low temperature limit, we expect that the bootstrap equation reduces to
\begin{equation}
\begin{aligned}
\ex{\fr{\beta}{2}\fr{c-1}{24}} \ex{\fr{\bar{\beta}}{2}\fr{c-1}{24}}
&=
\sum_{p}C_{ABp}^2  
\pa{\pa{\fr{\beta}{2\pi}}^{\fr{c-1}{4}-4(h_A+h_B)}\pa{16}^{h_p}
\ex{-\fr{2\pi^2}{\beta}\pa{h_p-\fr{c-1}{24}}} 
}\\
& \hspace{10ex} \times\Biggl( \pa{\fr{\bar{\beta}}{2\pi}}^{\fr{c-1}{4}-4(\bar{h}_A+\bar{h}_B)}\pa{16}^{\bar{h}_p}
 \ex{-\fr{2\pi^2}{\bar{\beta}}\pa{\bar{h}_p-\fr{c-1}{24}}} 
\Biggr)  .
\end{aligned}
\end{equation}
Moreover, if we assume $c>1$, then we can approximate the sum by an integral. At this stage, we can evaluate the average of the light-light-heavy OPE coefficients by using the inverse Laplace transformation
\begin{equation}
\overline{C_{ABp}^2}  \ar{h_p,\bar{h}_p \to \infty} 16^{-(h_p+\bar{h}_p)} \ex{-\pi\s{\fr{c-1}{6}\pa{h_p-\fr{c-1}{24}}}-\pi\s{\fr{c-1}{6}\pa{\bar{h}_p-\fr{c-1}{24}}}},
\end{equation}
where the average is over all primary operators of fixed dimensions $h_p, \bar{h}_p$.
The saddle point of the inverse Laplace transformation is
\begin{equation}
\fr{h_p}{\pa{\log \e}^2}\simeq \fr{c-1}{24},
\end{equation}
which is in the regime $ \fr{h_p}{(\log \e)^2} \nll 1$ and, consequently, this analysis is self-consistent.

Note that the high-low temperature limit bootstrap equation in other setups reveals the universal formulae of heavy-heavy-light \cite{Kraus2016, Hikida2018, Romero-Bermudez2018, Brehm2018} and heavy-heavy-heavy \cite{Cardy2017} OPE coefficients. It can be argued that all such formulae, including the above heavy-light-light OPE coefficients, are analogs of the Cardy formula \cite{Cardy1986a}. 
\section{Discussion}\label{sec:discussion}

In this paper, we give one analytic proof of our previous conjecture \cite{Kusuki2018, Kusuki2018b} for large $c$ conformal blocks by using the monodromy method.
In particular, we show that the asymptotic behavior of the large $c$ ABBA blocks drastically changes at $h_{A,B}=\fr{c}{32}$. Alternatively, we can also say that our analytic derivation of the asymptotic properties is well verified by numerical computations.

Our results suggest that, in the bulk, the behavior of a collision between two heavy particles can have an interesting transition associated with their masses.  Unfortunately, we do not have the tools with which we can handle such geometry. Nevertheless, the simplification of the calculation of the conformal blocks suggests that the limit corresponding to $x \to 1$ allows us to investigate the bulk case. This limit corresponds to a late time limit \cite{Fitzpatrick2014,Hijano2015a} both in the Euclidean and Lorentzian space. It means that a geometry with two heavy particles (or black holes) could be different based on whether their masses are larger or smaller than $\fr{c}{32}$. We believe that the equivalence between solving the monodromy method in the boundary side and solving the Einstein equation in the bulk side \cite{Hijano2015a,Alkalaev2016} could be a key to understanding the bulk interpretation of this transition.

In \cite{Chang2016}, one can see the transition at $h_{A,B}=\fr{c}{32}$ in the fusion transformation, which relates s-channel and t-channel of conformal blocks \cite{Ponsot2004,Ponsot1999,Teschner2001a} (also see \cite{Teschner2014}),
\begin{equation}
\begin{aligned}
\ca{F}^{21}_{34}(h_{\a_s}|x)=\int_{\bb{S}} \dd \a_t {\bold F}_{\a_s, \a_t} 
   \left[
    \begin{array}{cc}
    \a_2   & \a_1  \\
     \a_3  &   \a_4\\
    \end{array}
  \right]
  \ca{F}^{41}_{32}(h_{\a_t}|1-x),
\end{aligned}
\end{equation}
where the contour $\bb{S}$ runs from $\fr{Q}{2}$ to $\fr{Q}{2}+ i\infty$. When $\a_1+\a_4<\fr{1}{2}$ or $\a_2+\a_3<\fr{1}{2}$, the poles of the fusion kernel ${\bold F}_{\a_s, \a_t} $ cross the contour  $\bb{S}$ and, thereby, the contour $\bb{S}$ has to be deformed. 
In \cite{Chang2016}, only the special case where all external states have the same dimension is considered but we can straightforwardly generalize the discussion to the above statement using the work of \cite{Ponsot2004}.
We will show that this also leads to the same transition of the conformal blocks in the future \cite{Kusuki2019}.

In fact, it might be possible to show our results from this fusion transformation because when we take the limit $1-x =\e\to 0$ in the left hand side, then the contributions from the conformal blocks in the right hand side can be approximated as $\e^{h_{\a_t}-h_2-h_3}$.
In the case where $\a_1+\a_4<\fr{1}{2}$ or $\a_2+\a_3<\fr{1}{2}$, the dominant contribution comes from the pole of the fusion kernel and, consequently, the  dominant contribution can be expressed by our results presented in Section \ref{sec:block}. Otherwise, we obtain $\e^{\fr{c}{24}-h_2-h_3}$ by integration along the contour  $\bb{S}$. However, it is possible that we cannot estimate the regime of validity of this approximation straightforwardly from this perspective, since the fusion kernel is too complicated.
This approach is interesting because the fusion kernel has the bulk interpretation as suggested in \cite{Chang2016} and, therefore, it could shed light on the interpretation of the transition occurring in the bulk side.

This work can be thought of as an analytic proof of the statement that the general solution to the Zamolodchikov recursion relation has an asymptotic property like the Cardy formula. However, there is currently no proof of that statement directly from the Zamolodchikov recursion relation. Therefore, one interesting topic for future work is understanding the mechanism resulting in such asymptotic behavior, by examining the recursion relation itself. This challenge is important because the monodromy method cannot probe the non-perturbative contributions, which appear in the forms derived by the Zamolodchikov recursion relation. In all cases, it is also important that future work reveals the higher order, and non-perturbative corrections to the large $c$ conformal blocks. Moreover, there is another reason why the challenge is important. In \cite{Kusuki2018}, we found that another transition along lines $h_A, h_B=\fr{c}{32}$ for AABB blocks (NOT ABBA blocks). However, we cannot find this transition by using the monodromy method used in this paper. We expect that there is another mechanism causing this transition and  it can be clarified in the analytic study of the recursion relation.

\section*{Acknowledgments}

We are grateful to Tadashi Takayanagi for fruitful discussions and
comments. We also thank Henry Maxfield, Ryo Sato and Sylvain Ribault for useful conversations about the subject of this study, and in particular Yasuaki Hikida and Jared Kaplan for reading the draft of this paper and giving us valuable comments. 
YK is supported by JSPS fellowship.
We would like to thank Editage for English language editing.
In this research work, we used the supercomputer of ACCMS, Kyoto University.

\bibliographystyle{JHEP}
\bibliography{MM}

\end{document}